\documentstyle [11pt,epsf] {article}

\begin{document}
\begin{titlepage}
\begin{center}
\vspace{2cm}
\Large 
{\bf The Star Formation Histories of Local Galaxies: Continuous or Intermittent?}
\\                                                     
\vspace{0.7cm} 
\large
Guinevere Kauffmann$^{1}$, St\'{e}phane Charlot$^{2,1}$, Michael L. Balogh$^{3}$\\
\vspace{0.3cm}
\small
{\em $^1$Max-Planck Institut f\"{u}r Astrophysik, D-85748 Garching, Germany} \\
{\em $^2$ Institut d'Astrophysique du CNRS, 98 bis Boulevard Arago, F-75014 Paris, France} \\
{\em $^3$ Department of Physics, University of Durham, South Road, Durham DH1 3LE, UK} \\

\vspace{0.5cm}
\begin {abstract}
\large      
We explore a model in which efficient star formation in galaxies is triggered by
merging satellites. We show that the merger/interaction rates predicted by
hierarchical galaxy formation models depend strongly on galaxy mass.
If a merger of a  
satellite larger than 1\% the mass of the primary triggers efficient star formation,         
low-mass dwarf galaxies experience strong bursts separated
by quiescent periods lasting several Gyr. Massive galaxies, such as our own Milky Way,
are perturbed by a 1\% satellite every few hundred million years, and thus 
have fluctuating, but relatively continuous star formation histories.
We study the spectral signatures of a population of galaxies undergoing intermittent star formation 
by combining the models with the latest version of the  Bruzual \& Charlot  
spectral synthesis code. We concentrate on spectral indicators that are sensitive primarily
to stellar age, rather than to metallicity and extinction. 
We show that if a population undergoes intermittent rather than continuous    
star formation, the signatures of the bursts should be evident from the observed dispersion
in the star formation rates, H$\delta$ equivalent widths and gas mass fractions of   
galaxies with 4000 {\AA} break strengths indicative of recent or ongoing star formation.
\vspace{0.8cm}

Keywords: galaxies:formation--galaxies:interactions--galaxies:stellar content\\
--galaxies:starburst

\end {abstract}
\vspace{0.8cm}
\end{center}
\normalsize
\end {titlepage}

\section {Introduction}          

There is considerable evidence that the star formation rates in many galaxies
have not been monotonic with time, but have instead exhibited significant fluctuations.
Analyses of the colour-magnitude diagrams of stars in  Local
Group galaxies show that no two Local Group members have identical star formation
histories (see Tolstoy 2000, Grebel 2000 for recent reviews). The lowest-mass systems,
the dwarf spheroidals and the dwarf irregulars, appear to have formed 
stars in one or two discrete episodes or `bursts'. 
More massive dwarf  galaxies, such as the Large and Small Magellanic Clouds,
seem to have formed stars continuously, but at a
variable rate (Dolphin 2000; Olsen 2000). Recent studies have
indicated that star formation in our own Milky Way has also been variable.
Rocha-Pinto et al (2000) analyzed chromospheric age indicators in a sample of solar
neighbourhood stars
and found fluctuations in the star formation rate with
amplitudes greater than a factor of 2--3 on timescales of 0.2--1 Gyr.
These results were confirmed by an analysis of solar neighborhood data from the Hipparcos
catalogue by Hernandez, Valls-Gabaud \& Gilmore (2000).

It is not yet possible to derive star formation histories for galaxies 
outside the Local Group. Instead, measurements of colours and 
spectral features  provide a `snapshot' of the stellar populations and  
current star formation rates of these objects. 
Traditionally, galaxies are subdivided                 
according to morphology, luminosity, or local overdensity, and trends in spectral
properties or colour  are studied. Analyses of the
global star formation rates (SFRs) of galaxies have been carried out by
studying the H$\alpha$, ultraviolet continuum and far-infrared emission from galaxies (see Kennicutt
1998 for a review). Their relative star formation rates (normalized to unit mass
or luminosity) have been found to depend strongly on morphological type (Cohen 1976;
Kennicutt \& Kent 1983), on stellar mass (Gavazzi \& Scodeggio 1996; Gavazzi et al 1998)
and on environment (Kennicutt 1983; Loveday, Tresse \& Maddox 1999; Couch et al 2000).
Late-type galaxies, low-mass galaxies and galaxies in low-density regions
form stars at higher relative rates than massive, early-type galaxies in groups and clusters.
Only a few studies address the question of variability in star formation (Glazebrook 
et al 1998; Sullivan et al 2000). This is because the galaxy samples analyzed so far
have been small or inhomogeneously selected. This situation will change dramatically
with the next generation of large redshift surveys such as the Sloan
Digital Sky Survey (SDSS) (York et al 2000) and Two Degree Field (2dF) 
galaxy redshift survey (Folkes et al 1999), 
which will obtain colours and spectra of millions of 
galaxies over the next years.

The physical processes that control how rapidly a galaxy is able to 
convert its available gas into stars are not well understood.
One popular idea is that there are two basic `modes' of star
formation: 1) an underlying, low-efficiency, continuous mode 
that is present in all galaxies; and 2) a more efficient `burst' mode 
that is triggered when gas is compressed as a result of a merger or other
dynamical disturbance. 

Low surface brightness galaxies (LSBs) may be examples
of objects that have have formed all or most of their stars in the low-efficiency
mode.  Analyses of the stellar populations of LSBs 
indicate that these galaxies do not experience a delayed onset of
star formation as proposed by McGaugh \& Bothun (1994), 
but have formed their stars  slowly over
a Hubble time (van Zee et al 1997; van den
Hoek et al 1999; Matthews, Gallagher \& van Driel 1999). In addition LSBs have higher
gas mass fractions than normal spirals (De Blok et al 1996), further supporting the idea
that the efficiency of star formation is  lower in LSBs than in high surface brightness
spirals.  Finally, LSBs are also clustered
much more weakly than normal galaxies on scales less than 2 $h^{-1}$ Mpc,
indicating that they are not located in environments where interactions and mergers
are common (Mo, McGaugh \& Bothun 1994).

The ultra-luminous infrared galaxies (ULIRGs) are the most dramatic examples
of galaxies that form stars in the burst mode. Their star formation rates are
estimated at hundreds of solar masses per year and most of these objects 
have turned out to be interacting or merging  galaxies  
(Borne  et al 2000). 
Numerical simulations have demonstrated that a merger of two equal mass disk galaxies
is extremely effective at channelling gas to the centre of the remnant, where it
can fuel a nuclear starburst (Barnes \& Hernquist 1991). These mergers also result in the
formation of elliptical galaxies. 

Simulations have
also shown that the distribution of gas in disk galaxies may also 
be strongly perturbed during minor mergers.  Hernquist \& Mihos (1995)  
simulated mergers between gas-rich disks and companions a tenth
the mass of the primary, and found that the disk developed strong
spiral structure. The gas lost angular momentum as a consequence of the 
formation of strong shocks, and a significant fraction was driven
into the inner regions of the disk. Byrd \& Howard (1992) showed that
satellites larger than 1\% the mass the primary galaxy could excite large-scale
tidal arms in the disk of the primary, which were able to survive as long as 
six disk edge rotations. If minor mergers lead to gas compression in disk galaxies
and if stars form more efficiently in regions of higher gas density (Kennicutt 1999),
then these events may be responsible for the fluctuating star formation rates
observed in galaxies. 

In this paper, we explore a model in which efficient star formation is
triggered by mergers. We compute
galaxy merging rates as a function of  mass and 
of redshift using the predictions of hierarchical galaxy formation models
(Kauffmann, White \& Guiderdoni 1993; Kauffmann \& Charlot 1998).
A similar model for star formation  has been applied to the high-redshift  
galaxy population by
Somerville, Primack \& Faber (2001). Here we focus on the star formation histories
of present-day galaxies as a function of their luminosity or mass.
In \S2 and \S3, we show that low-mass galaxies are triggered much less frequently
than high-mass galaxies. As a result, efficient star formation occurs much more intermittently
in low-mass galaxies and a larger percentage of these  
objects will have low surface brightness and high gas-to-stellar mass  ratios.
In \S4, we explore the spectral signatures of a bursting population of galaxies
by combining our models with the latest version of the  Bruzual \& Charlot (1993) 
spectral synthesis code.               
We show that if a population of galaxies undergoes intermittent rather than continuous    
star formation, the signatures of the bursts should be apparent in the widths 
of the distributions of 
4000 {\AA} break strengths, Balmer absorption line equivalent widths, current
star formation rates and gas mass fractions.            

\section {Merging/triggering rates as a function of mass}

In hierarchical theories of structure formation, large structures such as galaxies, groups
and clusters form through the continuous aggregation of non-linear halos of
dark matter into larger and larger units.
Merging trees that track the merger paths of dark matter halos
as a function of time can be generated using the extended Press-Schechter
theory (Cole \& Lacey 1993; Kauffmann \& White 1993; Kolatt \& Somerville 1999;
Cole et al 2000) or directly from N-body simulations (Roukema et al 1997; Kauffmann
et al 1999; Van Kampen et al 1999).  

As discussed in Kauffmann \& White (1993), the merging histories of dark matter
halos depend both on cosmological parameters, such as $\Omega$, $\Lambda$ 
and $\sigma_8$, and on the mass of the halo. In particular, high-mass halos have    
formed through more late mergers  than low-mass halos.
This is illustrated in the left-hand panel of Fig.~1, where we plot the average time to the last
merger of an object with at least a tenth the mass of the primary halo as
a function of the circular velocity of the halo at the present day.
Here and in the rest of the paper we adopt a $\Lambda$CDM cosmology with             
$\Omega=0.3$, $\Lambda=0.7$, $\sigma_8=0.9$ and $H_0=75$ km s$^{-1}$ Mpc$^{-1}$.
The procedure for computing the merging histories is described in detail
in Kauffmann \& White (1993).
The right-hand panel of Fig.~1  shows the fraction of halos  that have
{\em not} experienced a merger with another halo greater than a tenth of their mass
for a significant fraction (two-thirds) of a Hubble time.
Note that below $V_c \sim 200$ km s$^{-1}$, the fraction of undisturbed halos 
rises steeply. 

\begin{figure}
\centerline{
\epsfxsize=9cm \epsfbox{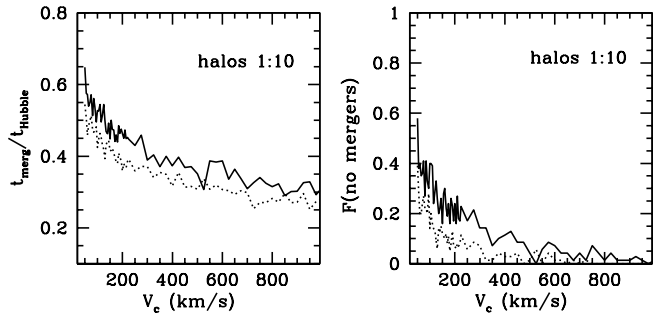}
}
\caption{\label{fig1}
\small
{\em Left:} The average time (in units of the Hubble time at time $t$) to the last 
merger with mass ratio greater than 1:10 
 as a function of the circular velocity of the halo. The solid line shows
results for present-day halos, the dotted line for halos at $z=1$.
{\em Right:} The fraction of halos that have not experienced at least a
1:10 merger for the past 2/3 $t_{Hubble}$.}  
\end {figure}
\normalsize

In hierarchical models, galaxies do not merge with each other at the same time
as their surrounding dark matter halos. N-body plus smoothed particle
hydrodynamics simulations show that as dark matter halos coalesce, the embedded
galaxies merge on a timescale that is consistent with dynamical friction estimates
based on their total (gas plus surrounding dark matter) mass (Navarro et al 1995).
In semi-analytic models of galaxy formation, the  halo merging trees and dynamical
friction-based estimates of galaxy merging rates are  
combined to estimate how frequently satellite galaxies merge with
central galaxies of given luminosity or stellar mass. 

\begin{figure}
\centerline{
\epsfxsize=9cm \epsfbox{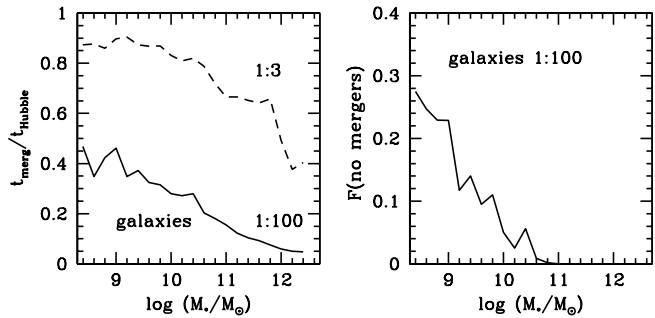}
}
\caption{\label{fig2}
\small
{\em Left:} The average time (in units of the Hubble time at time $t$) to the last 
merger with mass ratio greater than  1:100 (solid line)
and 1:3 (dashed line)  
as a function of the stellar  mass of the galaxy at the present day. 
{\em Right:} The fraction of galaxies that have not experienced at least a
1:100 merger for the past 2/3 $t_{\rm{Hubble}}$.}  
\end {figure}

\normalsize

In the left panel of Fig.~2, we plot the average time to the last merger of a satellite
with at least a third and a hundredth the mass of the central galaxy, as a function
of the stellar mass of the central galaxy at the present day.
As discussed in the introduction, 1:3 mergers are major events that trigger 
strong nuclear starbursts and lead to the formation of elliptical galaxies.
However, major mergers are rare. The majority of galaxies with present-day stellar
masses less than $\sim 10^{10} M_{\odot}$ never experienced a major merger.
Only galaxies with masses in excess of $\sim 10^{11} M_{\odot}$ have typically
undergone a major merger since redshift 1.  Minor mergers
are much more common. Milky-Way-type galaxies with masses of
$\sim 10^{11} M_{\odot}$  typically have a 1:100 merger every 500
million years. On the other hand, dwarf galaxies with masses $\sim 10^{8} M_{\odot}$ 
typically experienced their last 1\% merger more than 5 billion years ago.
The mass dependence of late-time mergers is illustrated again in the right
panel of  Fig.~2, 
where we plot the fraction of galaxies
that have not been disturbed by a 1:100 event for the past two-thirds of a Hubble time.
As can be seen, the fraction of unperturbed galaxies is a strong function
of galaxy mass, rising from near zero for $L_*$ galaxies to nearly a third for
$10^{8} M_{\odot}$ dwarfs.

In traditional models for the formation of disk galaxies, low surface brightness (LSB)
galaxies are assumed to form in halos with high angular momentum (Dalcanton, Spergel
\& Summers 1997; Mo, Mao \& White 1997). It is generally assumed that a fixed fraction
of the baryons in the halo cool to form the disk and that all disks have the same
mass-to-light ratio. This implies a relation between  the central surface brightness
of the disk and the mass of its halo, $\Sigma_0 \propto M_{halo}^{1/3}$. 
Because the angular momentum distribution of dark matter halos depends only weakly
on local overdensity (Lemson \& Kauffmann 1999),  
LSB galaxies are not expected to be clustered any differently from high surface 
brightness galaxies in such models.
This disagrees with direct observational evidence that
low surface brightness galaxies are much more weakly clustered on small scales
(Mo, McGaugh \& Bothun 1994). Another weakness of the models is that
they predict a luminosity-surface brightness relation that is much shallower 
than the one derived by Driver (1999) for galaxies in the Hubble Deep Field,
which has the form $\mu_e \propto 0.7 M_B$.

We suggest here that both problems may be solved in a scenario where
mergers trigger efficient star 
formation and increase the surface brightness of galactic disks.         
Because low-mass galaxies are triggered
less often, they will have larger gas fractions and higher
mass-to-light ratios  than high-mass galaxies 
(see \S4). In addition, LSBs would not survive in
dense environments where the probability of an encounter with another galaxy
is high.                   

In summary, merging rates 
depend strongly on mass in hierarchical cosmologies. Low mass galaxies
experience fewer recent mergers than high-mass galaxies. If mergers trigger
efficient  star formation, the star formation histories of 
dwarf galaxies should be characterized by much more widely-spaced
bursts than those of massive galaxies.
 
\section {A Toy Representation of Intermittent Star Formation} 

Detailed N-body plus hydrodynamic simulations are necessary in order to 
understand the triggering effects of  mergers in detail.
Mihos \& Hernquist (1995) have demonstrated that the nature of the gas inflows 
are sensitive not only to the orbital parameters of the infalling satellite,
but also to the structure of the primary galaxy, for example the presence
of a bulge. 
Detailed modelling of the effects of  mergers is beyond the
scope of this paper. Our goal is to study the spectral
signatures of a {\em population} of galaxies undergoing intermittent
rather than continuous star formation. To do this, we construct a toy model
of merger--triggered star formation. We appeal to
standard semi-analytic models of galaxy formation, which follow the formation
and evolution of galaxies within a merging hierarchy of dark matter halos.
These models include simple `recipes' that describe gas cooling, star formation,
supernova feedback and galaxy-galaxy merging (see for example Kauffmann et al 1999;
Somerville \& Primack 1999; Cole et al 2000).

Semi-analytic models have traditionally adopted a star formation law of
the form $M_* = \alpha M_{cold} /t_{dyn}$, where $M_{cold}$ is the mass of
cold gas in the disk, $t_{dyn}$ is the dynamical timescale of the galaxy, and
$\alpha$ is a free parameter controlling the efficiency of the conversion of cold gas
into stars. $M_{cold}$ is controlled by the cooling rate      
from the surrounding halo, by the rate  at which gas is turned into stars, and by the rate
at which supernovae reheat cold gas. The star formation efficiency
parameter $\alpha$ is tuned in order to match the $B$-band luminosity and the gas
mass of a Milky Way--type galaxy in a halo of circular velocity 220 km s$^{-1}$ at the
present day.
This star formation law results in a roughly continuous rate of star formation
in galaxy disks with a redshift--dependence that is 
similar for galaxies of different mass or circular velocity.
This is illustrated in  Fig.~3,  which also
shows that this star formation law results in very little scatter 
in the current star formation rates
of disk galaxies of fixed present-day circular velocity.
  
\begin{figure}
\centerline{
\epsfxsize=7cm \epsfbox{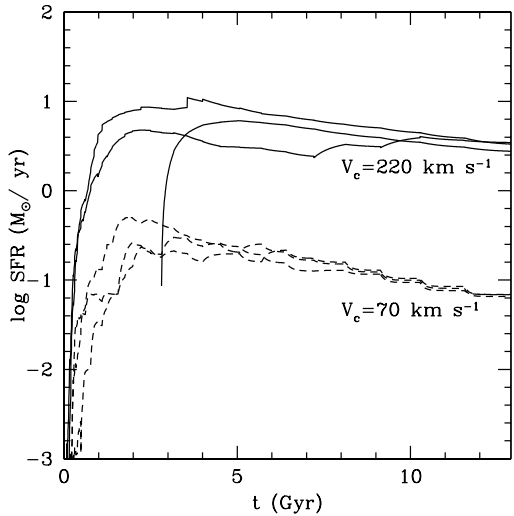}
}
\caption{\label{fig3}
\small
SFR histories of disks that form stars only in the quiescent mode. The upper curves show results
for disks forming in halos with present-day circular velocities of  220 km s$^{-1}$, 
while the lower curves show
results for disks with $V_c = 70$ km s$^{-1}$.}                                     
\end {figure}

One simple way to obtain fluctuating star formation histories in galaxies is to allow
the efficiency parameter $\alpha$ to vary if the galaxy experiences a merger.
In the semi-analytic models of Kauffmann et al (1999) and Cole et al (2000), this
only happens if a galaxy experiences a `major' merger with an object of
at least a third of its mass. In this case, all the available cold gas
is converted into stars on a timescale of $10^8$ years. However, as we have
seen, major mergers occur infrequently, particularly in low-mass galaxies.
In addition, they cannot explain fluctuations in star formation rates
on timescales of less than several Gyr in
high-mass galaxies. Recently, Somerville, Primack \& Faber (2001)
considered a model in which
every galaxy merger triggered a starburst with an efficiency
that scaled as a weak  power-law function of the mass-ratio of the merging galaxy pair.
The parameters controlling the burst efficiency were tuned to match the simulation
results of Mihos \& Hernquist (1994; 1996).

We will follow Somerville et al (2001) in assuming that
all mergers induce starbursts. However,
we choose to keep things conceptually simple by assuming that disks only form stars
in two `modes': a low-efficiency mode  ($\alpha_q$) 
and a high-efficiency mode ($\alpha_b$) that occurs
for duration $t_b$ after a merger with a satellite with mass
at least 1\% that of the primary galaxy.
Note that this is actually not very different to the prescription adopted by
Somerville at al (2000). According to their parametrization, the starburst 
efficiency depends
only weakly on the mass ratio: the efficiency of a burst triggered
by a 1:100 merger is only half that of a burst triggered by a 1:1 merger.

In our analysis, we explore the effect
of changing the relative efficiencies of the burst and quiescent modes by
varying $\alpha_b/\alpha_q$ (the absolute normalization of these parameters
is always chosen to obtain the same $B$-band luminosity for central galaxies in 
present-day halos of $V_c =$ 220 km s$^{-1}$). We also explore the effect of the 
the choice of  burst timescale $t_b$ on a variety of spectral features.   
Otherwise, the models are the same as in Kauffmann \& Charlot (1998) and
Kauffmann et al (1999), to which we refer for more details. 

Fig.~4 shows the star formation
histories of the central galaxies in three halos with $V_c =$ 220 km s$^{-1}$ 
at the present day (i.e., Milky Way-type galaxies). Here we have assumed
that the efficiency of the triggered mode is 10 times the efficiency
of the quiescent mode ($\alpha_b =10 \alpha_q$). We have further assumed
that the burst duration scales with dynamical time, $t_b\propto[0.7 + 
0.3(1+z)^3]^{-1/2}$, with $t_b=1$ Gyr at $z=0$. For comparison, Fig.~5 shows
the star formation histories of 3 central galaxies in halos of 70 km s$^{-1}$
(i.e., dwarf-type galaxies).

Because galaxies in 220 km s$^{-1}$ halos are triggered several times               
every Gyr on average, their star formation histories are continuous
over long periods. Occasional `gaps'  occur, but galaxies          
form stars in the efficient burst mode most of the time.                
`Dwarf' galaxies in 70 km s$^{-1}$ halos behave very differently. They experience
quiescent periods of many Gyr, punctuated by shorter bursts. 
At fixed cosmic time $t$, if there has been a long gap since the last episode of
efficient star formation,
the burst is stronger because the galaxy has had longer to accumulate more
cold gas. We also note that triggering is more frequent at early times. 

\begin{figure}
\centerline{
\epsfxsize=8cm \epsfbox{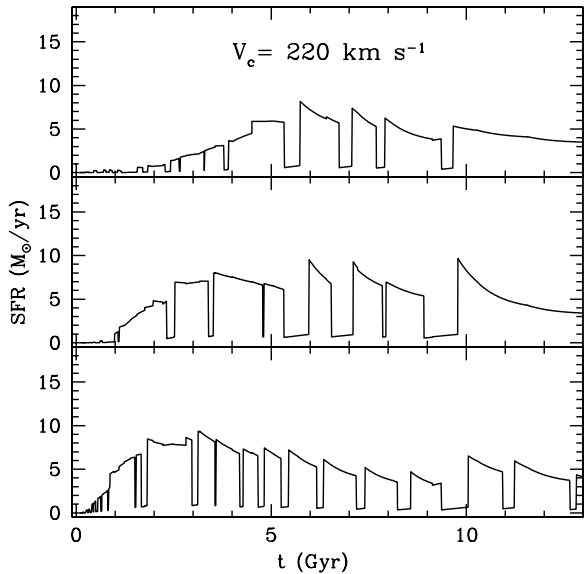}
}
\caption{\label{fig4}
\small
SFR histories of central galaxies residing in  halos with $V_c$= 220 km s$^{-1}$ at the present day
forming stars quiescently
and in a burst mode triggered by minor mergers. }                                     
\end {figure}

\begin{figure}
\centerline{
\epsfxsize=8cm \epsfbox{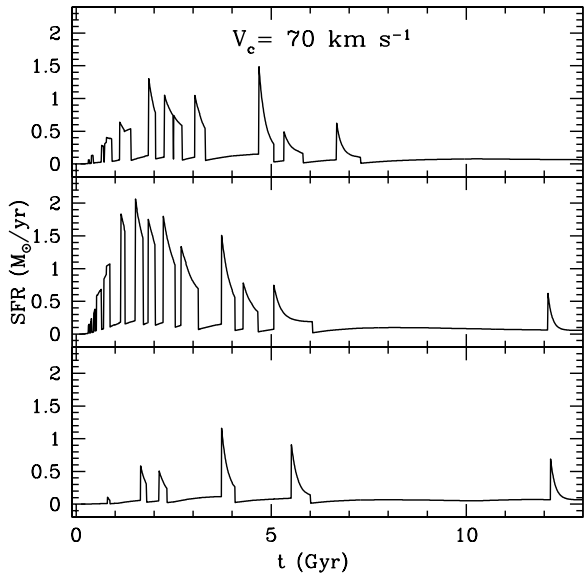}
}
\caption{\label{fig5}
\small
As in 4, but for central galaxies in halos with $V_c$= 70 km s$^{-1}$ today.}                                     
\end {figure}

The physical processes regulating 
star formation in satellite galaxies are likely to be different to the processes
that determine the star formation histories of central galaxies. 
In the semi-analytic models of Kauffmann et al (1999), the star formation rates
of satellites decline exponentially on a timescale of $\sim$ 1 Gyr because
these objects no longer accrete cold gas from the halo.
Ram-pressure stripping has been proposed as another mechanism that can rid a
galaxy of its gas and lead to the cessation of star formation in cluster
spirals (Gunn \& Gott 1972). This is not included in the models presented here.

We also assume here that satellite galaxies
within a halo can merge only with the central galaxy of that halo. Springel et al (2001)
used high-resolution N-body simulations to study the rate at which two 
satellite galaxies merge with each other
and found it to be  low (5\%) compared to that of
satellite--central galaxy mergers. `Non-sticking' collisions between satellite galaxies
are much more frequent (Tormen, Diaferio \& Syer 1998; Kolatt et al 2000) and these events
may also excite bursts of star formation (Moore et al 1996). In fact, there is
evidence that the last significant burst of star formation in the LMC may have
been caused by a tidal interaction with our Galaxy (Lin, Jones \& Klemola 1995).
Because we cannot follow galaxy orbits using  our semi-analytic 
approach, we are  not able to model the influence of collisions on
the star formation histories of individual satellite systems.       
Instead, we assume
that satellite galaxies form stars in the high-efficiency  mode for a time $t_b$ 
every time their parent dark matter halo merges with another more massive halo,
i.e. efficient star formation is triggered at the time of collapse of the
next level of the hierarchy. This is very similar to the
prescription adopted by Lacey \& Silk (1991) in their model of tidally
triggered galaxy formation.  
We will concentrate on the properties of star-forming field galaxies in this paper.
These are mostly the central objects of their halos, so the prescriptions adopted for star formation
in satellite systems will not be important for most
of our results.

We will explore four models with different 
ratios of burst--to--quiescent mode efficiency                                      
($\alpha_b/\alpha_q)$ and different burst timescales $t_b$. The properties of the models are
summarized in Table 1.                                    
Model Q is our `reference model' in which all star formation takes place in the quiescent mode
($\alpha_b=0$). 
In this model, central galaxies form stars at a rate that is regulated by the amount of
gas that cools from the surrounding dark matter halo. Satellite galaxies do not accrete any
gas, and as a result they always have lower star formation rates and redder colours than
the central galaxy in their halo.
In models A--C, central galaxies
form stars both quiescently  and in a burst mode, which is triggered by galaxy-galaxy mergers.
Merging of the parent halo with a more massive halo triggers star formation  in satellite 
galaxies, as described above.
Comparison of model B to model A  illustrates the effect of increasing        
$\alpha_b/\alpha_q$. Comparison of model C to model A
illustrates the effect of decreasing the burst timescale $t_b$.

In order to develop some intuition for how star formation occurs in these models, we
define 3 different `regimes' in which stars form:  
\begin {enumerate}
\item {\em Low efficiency}:  star formation that occurs in an unperturbed galaxy.
\item {\em High efficiency burst}: high-efficiency star formation 
  that is triggered by a merger, but that continues for less than 1 Gyr.              
\item {\em High efficiency continuous}: high-efficiency star formation 
  that is triggered by a series of mergers 
  and that continues for longer than 1 Gyr.
\end {enumerate}

In the left-hand panels of Fig.~6, we plot the average fraction of stars formed in the past 4 Gyr 
in each of these 3 regimes as a function of the luminosity of the galaxy. Results are 
shown for central galaxies in models A, B, and C (in model Q, the fraction of stars formed
in the low efficiency regime is unity at all luminosities).
In the right-hand panels, we
show the distribution of galaxies as a function of the mass fraction of stars
formed in the burst mode ($F_{burst}$). The hatched histograms are for galaxies of low
luminosity ($M_R \sim-18.0$), the shaded histograms are for galaxies of intermediate luminosity
($M_R\sim-20.5$) and the open histograms are for bright galaxies ($M_R \sim -23.5$).  

In all three models where efficient star formation is triggered by mergers, 
low-luminosity  galaxies form most of their stars in the low-efficiency regime and bright
galaxies in the high-efficiency continuous regime. The contribution of
bursts is greatest in galaxies of intermediate and high luminosities.
In model B, where $\alpha_b/\alpha_q$ is larger than in model A, the fraction of stars formed in
bursts increases and the fraction formed quiescently decreases, particularly
for galaxies with low and intermediate luminosities. Unlike model A, Model B has an extended 
tail of low-luminosity  galaxies that have formed most of their stars in the burst
mode.  In model C, where the burst timescale
is shorter than in model A, a larger  fraction of stars in bright galaxies
form in the low-efficiency regime. In \S4, we show how these differences between the three
models translate into differences in the distributions of strengths of spectral features such
as the H$\delta$ equivalent width.

\begin{figure}
\centerline{
\epsfxsize=12cm \epsfbox{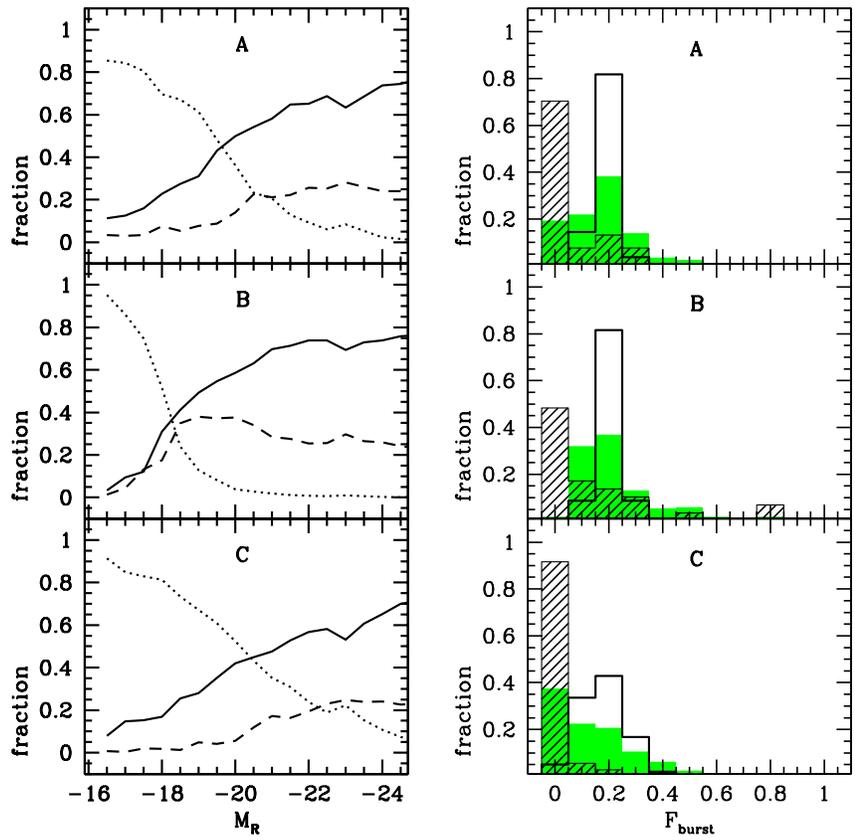}
}
\caption{\label{fig6}
\small
Left: The average fraction of stars formed in the past 4 Gyr 
in the low-efficiency (dotted), burst (dashed) and high-efficiency continuous (solid)
regimes as a function of the luminosity of the galaxy.  
Right: the distribution of galaxies as a function of the fraction of stars
formed in the burst mode ($F_{\rm burst}$. The hatched histograms are for galaxies of low
luminosity ($-18.5 \leq M_R < -17.5$), the shaded histograms are for galaxies of 
intermediate luminosity ($-21.0 \leq M_R < -20.0$) and the open histograms are for bright 
($-24.0 \leq M_R < -23.0$) galaxies.}  
\end {figure}

\section{Spectral Signatures of Bursts}

\subsection {Spectral Indicators}
In this section, we introduce a number of spectral indicators that are useful diagnostics
of the star formation histories of galaxies. Our choice of indicators is motivated by
the following considerations:
\begin {enumerate}
\item Sensitivity of the indicator to stellar age, rather than to metallicity and 
  dust extinction.
\item Existence of measurements of the indicator for magnitude-limited
      samples of galaxies. 
\item Ability of models to predict the behaviour of the index as 
      a function of stellar age.
\end {enumerate}

\subsubsection {The 4000 Angstrom Break: $D(4000)$} 
Discontinuities in the spectra of galaxies are produced
by the accumulation of a large number of spectral lines in a narrow wavelength region,
which produces a sharp opacity edge reflected as a break
in the spectrum. The main contribution to the opacity comes from ionized metals. 
In hot stars, the elements are multiply ionized and the opacity decreases at optical wavelengths,
so the optical breaks are small for young stellar populations and large for old, metal-rich
galaxies.

The break occurring at 4000 {\AA} is the strongest discontinuity in the spectrum
and the one for which the most galaxy data are available.
A break index $D(4000)$ has been defined by Bruzual (1983) as the ratio of the average flux 
density F$_{\nu}$ in the bands 4050--4250 {\AA} and 3750--3950 {\AA}. 
A definition using a narrower continuum band has recently been adopted by         
Balogh et al (1999). The principle advantage of the narrow definition is that the index
is then less sensitive to reddening effects.

Fig.~7 illustrates the accuracy of the $D(4000)$ index as a stellar age indicator in
comparison to broadband colours. We have used the latest version of the
Bruzual \& Charlot (1993)  population synthesis models 
 to compute the value of the index 
for galaxies with a wide range of different star formation histories.
Galaxies are assumed to form stars according to the law SFR$(t) \propto \exp [ \gamma t(\rm{Gyr})]$
from redshift $z_{form}$ to the present day. We pick $z_{form}$ randomly over
the interval 0.2--10 and $\gamma$ over the interval $-2$--$+1$ (this allows the SFR to either
decrease or increase with time).
In order to obtain as wide a range of SFR histories as possible,
we also superimpose random `bursts'  with amplitude 2--100 times the
underlying star formation rate and with duration  0.1--1 Gyr.   
In Fig.~7, we plot the value of the $D(4000)$ index versus the 
$V$-light weighted mean stellar age of the galaxies.
Results are shown for two metallicities: solar and 20\% solar.
Here and in the rest of the paper, we have adopted a Kennicutt (1983b) initial mass function
with upper and lower mass cutoffs at 0.1 and 100 $M_{\odot}$. 

There is an excellent, metallicity-independent correlation between $D(4000)$ and 
$V$-light weighted age for galaxies with
mean stellar ages less than $\sim 2$ Gyr and $D(4000) < 1.6$. 
The $D(4000)$ index depends more strongly  on metallicity
at ages older than a few Gyr. In contrast, optical-infrared colours such as $V-K$ depend
on metallicity at all ages. This is because the infrared emission is always dominated by giant
stars (supergiants, AGBs or RGBs) with temperatures that are sensitive to              
metallicity. The $B-V$ colour is a much better age indicator. Its main drawback
is the uncertainty introduced by dust. Wang \& Heckman have shown that the
mean $B$-band optical depth of a $L_*$ spiral galaxy is $\tau_B \sim 0.8$.
This implies that the typical correction to the $B-V$ colour from dust is $\sim 0.17$,
implying a factor $\sim 10$ uncertainty in derived age for blue, star-forming galaxies.
This amount of extinction introduces an 8\% correction to $D(4000)$, corresponding
to a factor $\sim 3$ uncertainty in mean age. 
This can be further improved by adopting the narrower definition of the break used 
by Balogh et al (1999), for which the extinction correction is only $\sim$ 3.4\%.  

Fig.~7 shows that the dynamic range in $D(4000)$ at ages less than a few Gyr is small. However, 
the index can easily be measured with errors of only a few hundredths.
Emission from the H-recombination continuum blueward of the break is negligible
except for stars with ages less than $3 \times 10^6$ years, so we expect that
the model predictions will be accurate for all but the most strongly star-forming galaxies.

\begin{figure}
\centerline{
\epsfxsize=11cm \epsfbox{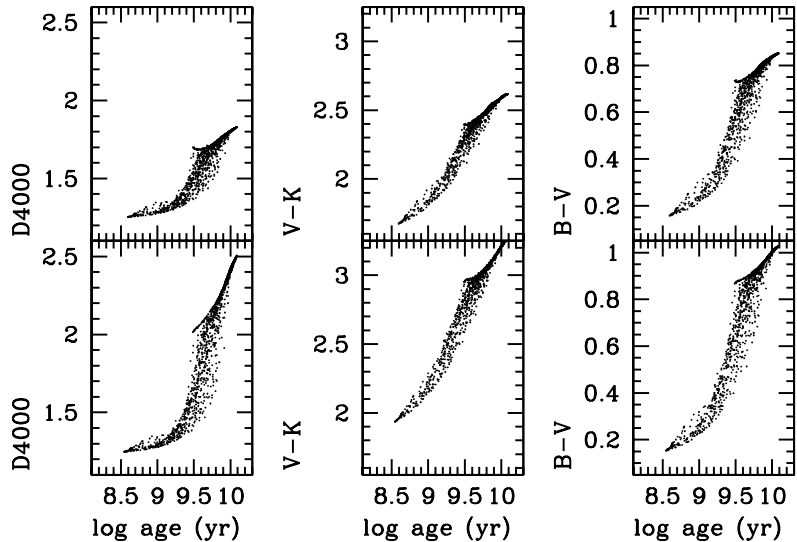}
}
\caption{\label{fig7}
\small
$D(4000)$, $V-K$ and $B-V$ as a function of $V$-light weighted mean stellar age for galaxies
with a wide range of different star formation histories. The upper panel shows results
obtained at 20\% solar metallicity, while the lower panel is for solar metallicity. }
\end {figure}

\subsubsection { The post-burst indicator $H\delta_A$}
Strong H$\delta$ absorption lines arise in galaxies that have experienced
an intense burst of star formation that ended 1--2 Gyr ago. The peak occurs once
hot O and B stars, which have weak intrinsic absorption, have terminated their
evolution,
and the optical light from the galaxy is dominated by late-B to early F-type stars.
The spectral resolution of the Bruzual \& Charlot models is too low to measure
the H$\delta$ absorption feature reliably. We adopt the standard 
procedure of  parameterizing absorption-line
strengths as functions of stellar effective temperature, gravity, and metallicity 
(e.g., Worthey et al. 1994). Worthey \& Ottaviani (1997) derive two alternative
parameterizations of the $H\delta$ absorption feature using the Lick/IDS stellar
spectral library. These correspond to narrow (H$\delta_F$) and wide 
(H$\delta_A$) definitions of the central bandpass bracketed by two 
pseudo-continuum bandpasses. Since the H$\delta_A$ index is closer in 
definition to the index used in most observational analyses, we adopt it 
as our standard in this paper. We compute the index strengths of the galaxies        
in our model by weighting the contributions from individual
stars by their level of continuum (see, e.g., Bressan, Chiosi \& Tantalo 1996
for a description of this standard procedure).

We also account for the contamination of stellar H$\delta$ absorption by 
nebular emission in galaxies containing young massive stars. We adopt
case-B recombination to compute the H$\delta$ emission produced by the 
ionizing radiation from these stars. If emission lines were attenuated 
by dust in the same way as continuum radiation, the emission correction
to the equivalent width would not depend on extinction.                                          
However, observations of nearby starburst galaxies indicate that the 
attenuation inferred from the ratio of H$\alpha$ to H$\beta$ is often higher than that
inferred from the spectral continuum (e.g., Fanelli, O'Connell \& Thuan 
1988; Calzetti, Kinney \& Storchi-Bergmann 1994). The most likely reason
for this apparent discrepancy is the finite lifetime of the dense clouds 
in which stars form, which causes the effective $B$-band absorption optical
depth of the dust to be typically a few times larger for emission lines 
than for the continuum radiation (Charlot \& Fall 2000). For $\tau_B=0.8$,
this implies a difference of at least a factor of 2 in the attenuation of
H$\delta$ line and continuum photons. We account for this in the models by 
reducing our dust-free
H$\delta$ emission corrections by a factor of two. Because the H$\delta$ emission
line is weak (compared to H$\alpha$ or H$\beta$), the actual emission correction
makes only a small difference to our results. Changing the value of the emission
correction can shift the values of
H$\delta$ reached  by galaxies with ongoing star formation by $\sim 1$ {\AA}, but
does not affect the values attained by galaxies once they have ceased forming stars.

The time evolution of H$\delta_A$ following an instantaneous burst of star
formation is shown in  Fig.~8. The solid line shows
the result for solar metallicity and the dashed line for 20\% solar metallicity.
We adopt the convention of positive index values for absorption and negative
values for emission. As can be seen, the dependence of the H$\delta_A$
index on metallicity is very weak. Note that H$\delta_A$ appears to be in `emission' at
late ages because the pseudo-continuum bandpasses surrounding the line are
depressed by strong metallic absorption lines (see  Fig.~6 of Worthey \&
Ottaviani 1997). 

In Fig.~9, we show the correlation between the H$\delta_A$ index and the quantity
$F_{burst}$ (defined as a the fraction of stars formed in  bursts over the 
past 4 Gyr; see \S3) for central galaxies of all luminosities in models A, B and C.
There is a reasonably good correlation between $F_{burst}$ and H$\delta_A$,
albeit with large scatter. The scatter arises because 
the value of the H$\delta_A$ index depends both on the mass of stars formed
in the burst and time at which the burst occurred. Nonetheless, it is still
possible to extract some information about the past star formation history
of an individual galaxy from the value of its H$\delta_A$ index.
Fig.~9 shows that galaxies that have formed stars continuously
over the past 4 Gyr (i.e. $F_{burst}=0$) have H$\delta_A$ equivalent widths
in the range $-2$--$+3$.
For  galaxies with  H$\delta_A > 4$ {\AA},  a significant ($> {\rm 15-20}$ \%) 
fraction of recent  star formation {\em must} have occurred in a burst.                  
For galaxies  with  H$\delta_A > {\rm 6-7}$ {\AA},  most   ($> {\rm 40-50}$ \%) 
recent star formation occurred in burst(s).
Note that the reverse does not hold. Just because a galaxy is observed with                
H$\delta_A < 4$ {\AA} today,  does not mean it experienced only continuous
star formation over the last 4 Gyr. We will show in \S5 that the index analysis
becomes more powerful if one considers the distributions predicted          
for a {\em population} of galaxies with similar star formation histories.

\begin{figure}
\centerline{
\epsfxsize=8cm \epsfbox{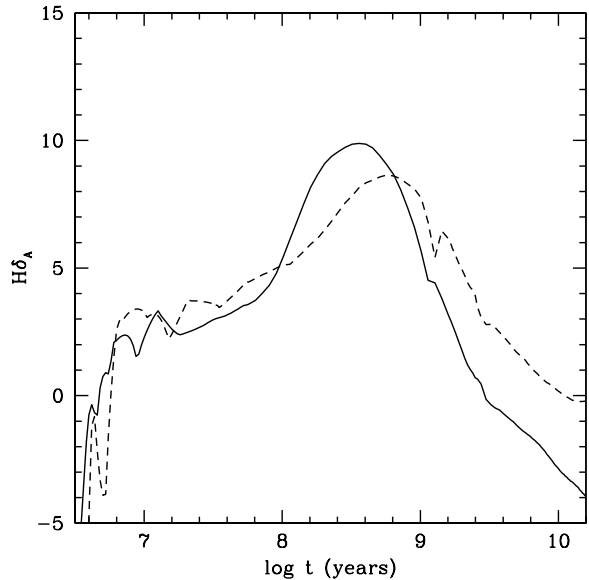}
}
\caption{\label{fig8}
\small
The H$\delta_A$ index is plotted against time following an instantaneous burst 
of star formation. The solid line is for solar metallicity and the dashed line
is for 20\% solar.} 
\end {figure}

\begin{figure}
\centerline{
\epsfxsize=8cm \epsfbox{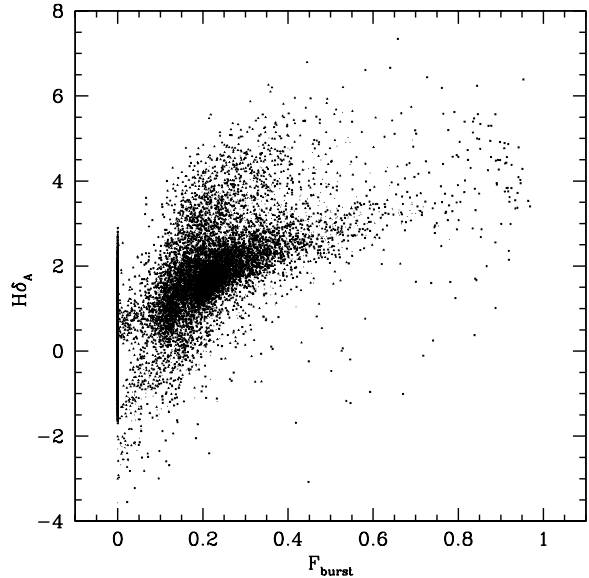}
}
\caption{\label{fig9}
\small
The H$\delta_A$ index is plotted against $F_{burst}$, the fraction of stars formed
in bursts during the past 4 Gyr for central galaxies in models A, B and C.
Note that the two ``sequences'' visible on the plot are simply
a consequence of the particular set of burst models we adopted. As discussed
later,  different models
lead to different density distribution of points in the diagram.}
\end {figure}

\subsubsection {Emission line indicators} 
Hydrogen recombination lines, such as H$\alpha$,
provide a measure of the {\em current} star formation rate in galaxies. The conversion factor beween
emission-line luminosities
and star formation rate is usually computed using an evolutionary synthesis
model. Only stars with masses greater than about 10 M$_{\odot}$ and lifetimes less than
30 Myr contribute significantly to the ionizing flux, so the recombination lines
provide a nearly instantaneous measure of the star formation rate,
independent of the previous star formation history of the galaxy. 
Extinction is the most important source of error in H$\alpha$-derived star formation rates.
Using integrated H$\alpha$ and radio fluxes of a sample of nearby galaxies,
Kennicutt (1983c) derives
a mean extinction A(H$\alpha$) = 0.8--1 mag.

The [OII]$\lambda$3727 forbidden-line doublet is also sometimes used as a star formation rate
indicator. It has been calibrated empirically (through H$\alpha$) as a quantitative SFR tracer
(Gallagher, Hunter \& Bushouse 1989; Kennicutt 1992). The main advantage of [OII] is that it can be 
observed in the visible out to redshifts
$z \sim 1.6$, and it has been measured in several large surveys of faint galaxies
(e.g. Cowie et al 1996; Ellis 1997). The main disadvantage is that the luminosities
of collisionally-excited lines are not directly coupled to the ionizing luminosity, and their
excitation is sensitive to metallicity and to the ionization state of the gas.
Recently, Jansen, Franx \& Fabricant  (2000) have found that the relation between
H$\alpha$ and [OII] depends strongly on magnitude. They attribute this
to systematic variations in reddening and metallicity as a function of galaxy luminosity.

Charlot \& Longhetti (2001) have quantified the errors in SFR estimates based
on optical emission lines by exploring a comprehensive set of models that
reproduced the observed properties of nearby galaxies. They showed that the H$\alpha$
and [OII] luminosities produced per unit SFR can vary by more than an order of
magnitude depending on the parameters of the ionized gas (metallicity, dust-to-heavy
element ratio, ionization state) and on the absorption by dust in the neutral 
interstellar medium. The large uncertainties in the derived SFR 
can be reduced to a factor of
only 2--3 when comparing the H$\alpha$ or [OII] emission to that of other lines 
such as H$\beta$, [OIII], [NII], and [SII]. Such detailed spectral information is
not currently available for large samples of galaxies, but this situation will change after
completion of future surveys such as SDSS and 2dF. 

\section {Burst Signatures in the Models}

In this section, we explore different ways of evaluating whether a population of galaxies   
observed at a single redshift has been forming stars continuously or intermittently.               
We generate catalogues of galaxies 
complete to $M_R = -16$ for models Q, A, B and C.
To achieve this, we first compute many realizations of the galaxy populations
in halos of different masses using the techniques described
in \S3. We then obtain the number density of galaxies 
of given luminosity, colour, star formation rate and spectral index by 
convolving these realizations with the abundance of halos calculated using
the Press-Schechter (1974) theory. 

We first study {\em global} 
distributions of spectral indicators, such as $D(4000)$ and H$\delta_A$.
In Fig.~10, we plot the distribution of 4000 {\AA} break
strengths as a function of absolute magnitude for 
models Q, A, B, and C.
The solid circles indicate the median value of $D(4000)$, while solid and dotted error
bars indicate the 25th--75th and 10th--90th  percentiles of the distributions.
In the pure quiescent model Q, the distribution of break strengths of faint galaxies 
extends to larger values than it does for bright galaxies.                
This is because a large fraction of the faint objects
are satellite galaxies that have run out of gas and that have little ongoing star formation,
whereas most bright objects are central galaxies which accrete gas from the surrounding halo
and form stars at a roughly constant rate (Fig.~3).
The break distributions of  bright galaxies exhibit significantly more scatter in models    
A, B and C than in model Q.
The break strengths of faint galaxies are larger in models where bursts
are important. As noted in \S2, faint galaxies are triggered infrequently
at late times, so they will have larger mean stellar ages in the burst models
than in the quiescent model.

\begin{figure}
\centerline{
\epsfxsize=8.5cm \epsfbox{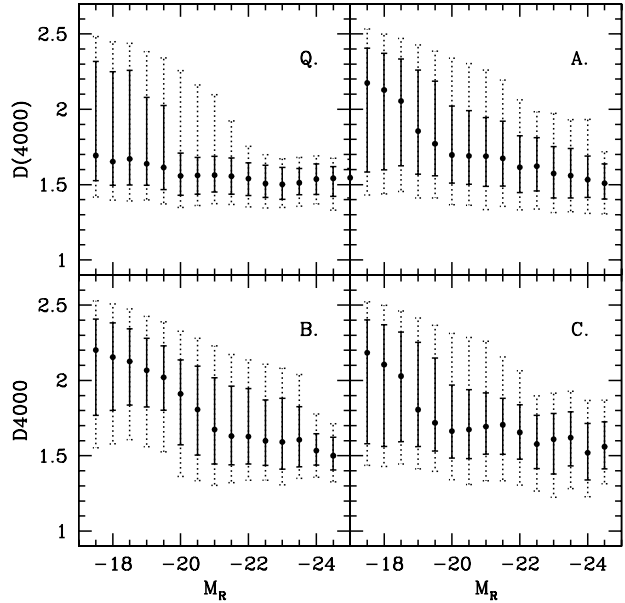}
}
\caption{\label{fig10}
\small
The $D(4000)$ index is plotted as a function of the $R$-band absolute magnitude of the galaxy
for the four models listed in Table 1. The solid circles indicate the median of the
distribution. The solid and dotted errorbars indicate the 25th--75th and 10th--90th
percentiles of the distribution.}
\end {figure}

As discussed in \S3, the physical processes regulating star formation in satellite
galaxies are likely to be different to  the processes that determine the star formation
histories of central galaxies. Ideally, one would like to
separate satellite from central galaxies when analyzing the  spectral index
distributions of galaxy populations. This can be done if one has environmental information about
the galaxies in the survey. Alternatively, one can use
the $D(4000)$ index strengths to select only those galaxies 
with young stellar populations. In the models, these are mostly central galaxies.

We showed in \S4 that the $D(4000)$ index is a good
mean stellar age indicator up to ages of a few Gyr and corresponding break strengths of 1.5--1.6.
In Fig.~11, we plot the H$\delta_A$ distribution as a function of absolute magnitude for
galaxies with $D(4000) < 1.4$ and $1.4 < D(4000) < 1.6$ for the same models as in Fig.~10.
In model Q,
these cuts select  actively star-forming central galaxies.  Because the star formation
in this model is determined only by cooling rates in dark matter halos, the H$\delta_A$ 
distributions exhibit very little scatter and do not depend on
galaxy luminosity. The H$\delta_A$ distributions of the burst models are very different.
In model A, galaxies with $M_R\sim -20$ show the strongest dispersion in H$\delta$
absorption. Fainter galaxies are triggered infrequently, so star-forming galaxies are seen mainly  
in the quiescent mode. Brighter galaxies are triggered very frequently, so individual bursts
overlap and the resulting star formation history is once again quasi-continuous (see Fig.~6).
If the burst timescale $t_b$ is shortened (model C), bright galaxies  also exhibit strong
H$\delta$ absorption. If the star formation efficiency in the quiescent model is very
small (model B), the fraction of faint galaxies undergoing strong bursts increases, and this 
is reflected in the fraction of H$\delta$-strong galaxies at magnitudes below $M_R = -20$.
In all the models, galaxies with the strongest H$\delta$ absorption have break values
in the range $1.4 < D(4000) < 1.6$. At smaller break strenghs, the H$\delta$
line is increasingly seen in emission rather than absorption.

\begin{figure}
\centerline{
\epsfxsize=10cm \epsfbox{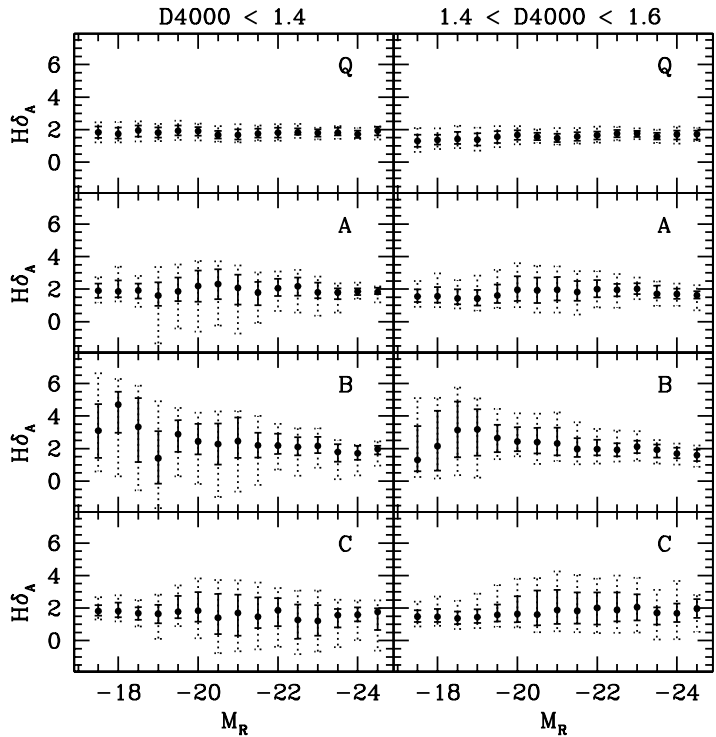}
}
\caption{\label{fig11}
\small
The H$\delta_A$ index is plotted as a function of the $R$-band absolute magnitude of the galaxy
for galaxies with $D(4000)<$1.4 and $1.4<D(4000)<1.6$.}
\end {figure}

There is currently  only one data set containing measurements of 4000 {\AA} break
strengths and H$\delta$ equivalent widths for  a large sample of local galaxies.
This is the Las Campanas Redshift Survey (LCRS; Schectman et al 1992,1996), which consists of 23700 
galaxy spectra with a mean redshift of 0.1 in six slices of 1.5 $\times$ 80 degrees. 
The galaxies are selected to have Kron-Cousins R-band isophotal magnitudes between
$15 < m_R < 17.7$ and surface brightnesses within the 3.5" fiber aperture
of $\mu_R < 21$ mag arcsec$^{-2}$. Zabludoff et al (1996) have developed an automated 
procedure for measuring the equivalent widths  of [OII]$\lambda$3727, Balmer lines (H$\delta$,
H$\gamma$ and H$\beta$) and the 4000 {\AA} break. A catalogue of these measurements
was kindly made available to us by A.Zabludoff.  

In Fig.~12, we plot the distribution of H$\delta$ equivalent widths of galaxies in the 
LCRS sample as a function of  R-band absolute magnitude and compare these
with the distributions given by models A, B and C. In both the data and the models, we have selected
galaxies with $D(4000) <  1.6$. The mean error on the EW(H$\delta$) measurements in the LCRS
is 0.85 {\AA}, with strong dependence on the strength of the feature but not on the 
absolute magnitude of the galaxy.
In our analysis, we have added errors to the values of H$\delta_A$ calculated in the models
by drawing from the distribution of errors as a function of H$\delta$ equivalent width
in the LCRS sample.  
We note that Zabludoff et al (1996) did not measure the H$\delta_A$ index, but
adopted a different procedure for estimating the equivalent width,
which involved finding the line center
and integrating outwards over                                  
100 pixels on either side of the line, where the continuum
was assumed to be reached. This procedure should in general lead
to larger measured equivalent widths, but unfortunately we are unable to 
quantify how much difference this would make in practice. In future work,
it will be very important to adopt consistent index 
definitions when comparing observations with theory.

Nevertheless it is interesting that in the LCRS the distribution of EW(H$\delta$) does appear to depend
on galaxy luminosity. 
The width of the distribution of  EW(H$\delta$) increases towards fainter magnitudes, reaching an
apparent ``peak'' for  galaxies with $M_R \sim -19.5$. The observations are in strong
disagreement with the pure quiescent model Q, where                              
the H$\delta_A$ distribution does not depend on  galaxy luminosity and where all galaxies have
H$\delta_A < 4$ {\AA}. However, it is probably premature to attempt to use the data to distinguish
between the various burst models.
The number of faint galaxies in the LCRS survey is  small, so 
the decrease in H$\delta$-strong galaxies at fainter magnitudes 
is not very significant. 
In addition, because the uncertanties in the measurements of EW(H$\delta$) are of the same
order or larger than the variations as a function of luminosity, a large sample of galaxies
with uncertainties less than $\sim 0.5 {\AA}$ would be extremely useful for this type of analysis.

\begin{figure}
\centerline{
\epsfxsize=8cm \epsfbox{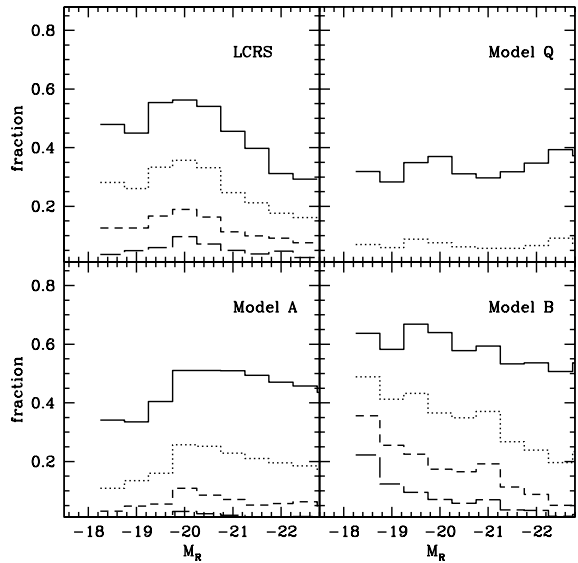}
}
\caption{\label{fig12}
\small
{\em Left:} The fraction of galaxies with $D(4000) < 1.6$ and H$\delta$ equivalent width greater
tha 2 {\AA} {solid}, 3 {\AA} (dotted), 4 {\AA} (short-dashed) and 5 {\AA} (long-dashed)
as a function of R-magnitude 
in the LCRS survey and in models Q,A and B. In the LCRS sample , reading from left to right
(i.e  from faint magnitudes to bright magnitudes), 
there are a total of 167, 349, 708, 1266, 1996, 2261, 1864, 897
and 198 galaxies in each magnitude bin.}   
\end {figure}

We now turn to an analysis of the information that could be provided by the distributions
of emission line strengths.
In Fig.~13,  we plot the instantaneous star formation 
rates of galaxies in the models normalized
by their $B$-band luminosity. Once again the distributions in model Q exhibit little scatter
and are independent of luminosity.
At small break strengths ($D(4000) < 1.4$)  models A,B and C have ``tails'' 
of bursting objects that form stars at many times the mean rate.
These tails are more pronounced for galaxies of intermediate 
luminosity, which are the most likely to be forming stars in the burst rather than the
low efficiency or high efficiency continuous regime at the present day (Fig.~6).
At intermediate break strengths  ($1.4 < D(4000) < 1.6$) the burst models are 
characterized by a shift towards star formation rates {\em lower} than
the mean. This is because many of the galaxies in the sample are caught just after the 
peak of the burst, when the massive
stars producing the ionizing radiation have disappeared, but the break strengths
are still quite small. This shift is strongest for bright galaxies in model C, which 
experience frequent bursts of short duration.

We conclude that although emission lines cannot be used as burst indicators in individual 
galaxies, in the analysis of galaxy populations they can serve as useful diagnostics of the past history
of star formation. Note, however, that the derived star formation rates must be 
accurate for this kind of analysis to be useful and that star formation rates
based on standard conversions using only [OII] or H$\alpha$, which have a factor of 10 or more
error (Charlot \& Longhetti 2001),  will not
suffice.

\begin{figure}
\centerline{
\epsfxsize=10cm \epsfbox{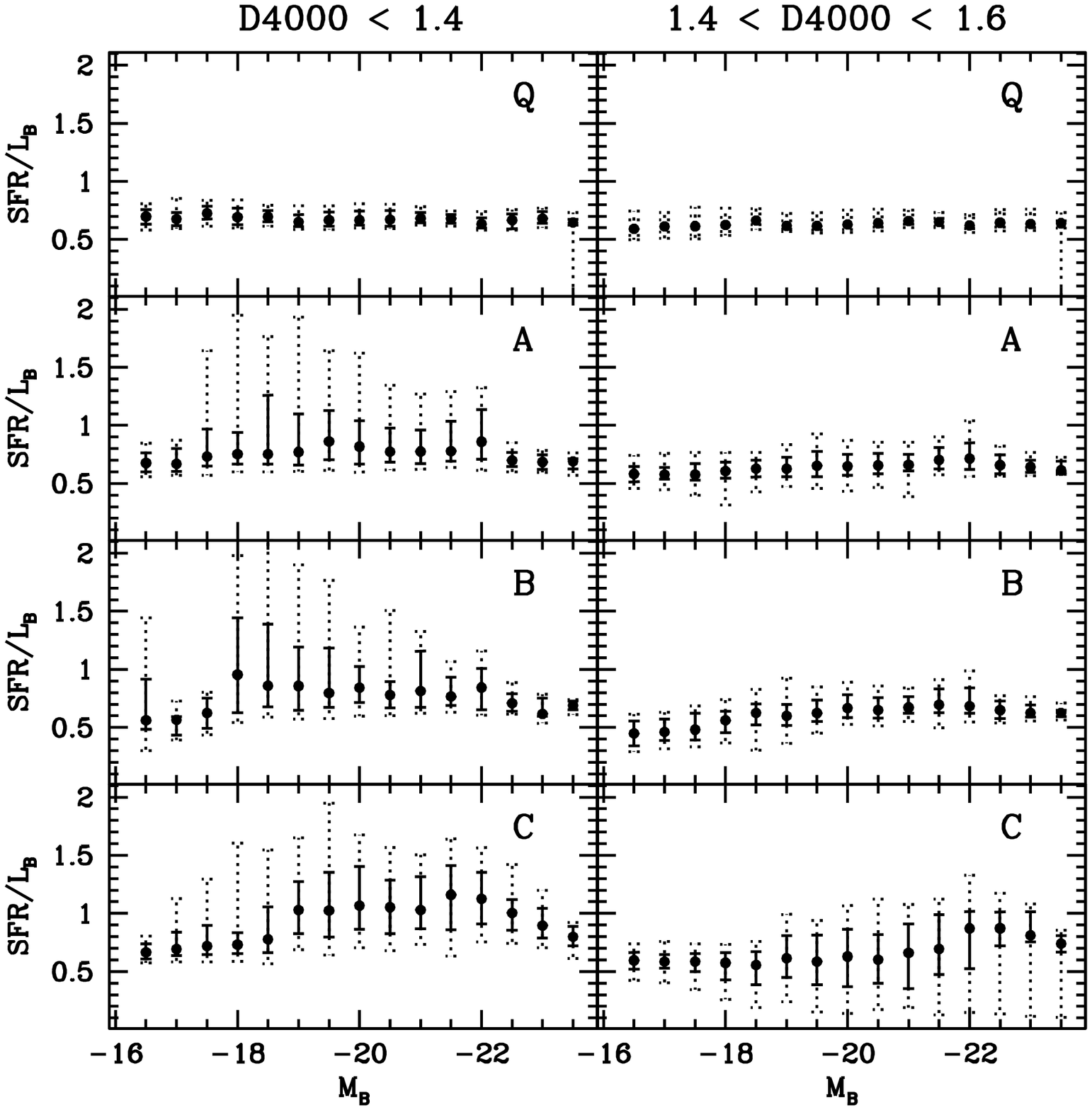}
}
\caption{\label{fig13}
\small
The star formation rates of galaxies normalized by their B-band luminosities
(in units of 1 M$_{\odot}$ yr$^{-1}$/ 10$^{10}$ $L_{B \odot}$)
are plotted as a function of  B-band absolute magnitude 
for galaxies with $D(4000)<1.4$ and $1.4<D(4000)<1.6$.}
\end {figure}

Important complementary information on the variability of star formation in galaxies
may be obtained by correlating the
stellar properties of galaxies with their gas masses. Although very little data currently
exist, the combination of the HI Parkes All-Sky
Survey of extragalactic HI (Barnes et al 2000)  and deep optical and  near-infrared imaging 
should produce strong constraints on burst models.
This is illustrated in Fig.~14, where we plot the distribution of $M_{gas}/M_{stars}$
for galaxies with $D(4000)< 1.6$. Once again, the distributions of $M_{gas}/M_{stars}$ 
in model Q do not depend on galaxy
luminosity and exhibit very little scatter. In the burst models,
faint galaxies are more gas rich than bright galaxies and also exhibit  greater
dispersion in gas fraction. Models with long burst timescales (models A and B) produce 
tails of faint, gas-poor galaxies with small break strengths. Such objects are missing 
in model C, which has a shorter burst timescale.
If the quiescent mode is extremely inefficient (model B),
there will be a large number of extremely gas-rich, but optically 
faint galaxies that should be easily
detected by surveys such as HIPASS.

\begin{figure}
\centerline{
\epsfxsize=8cm \epsfbox{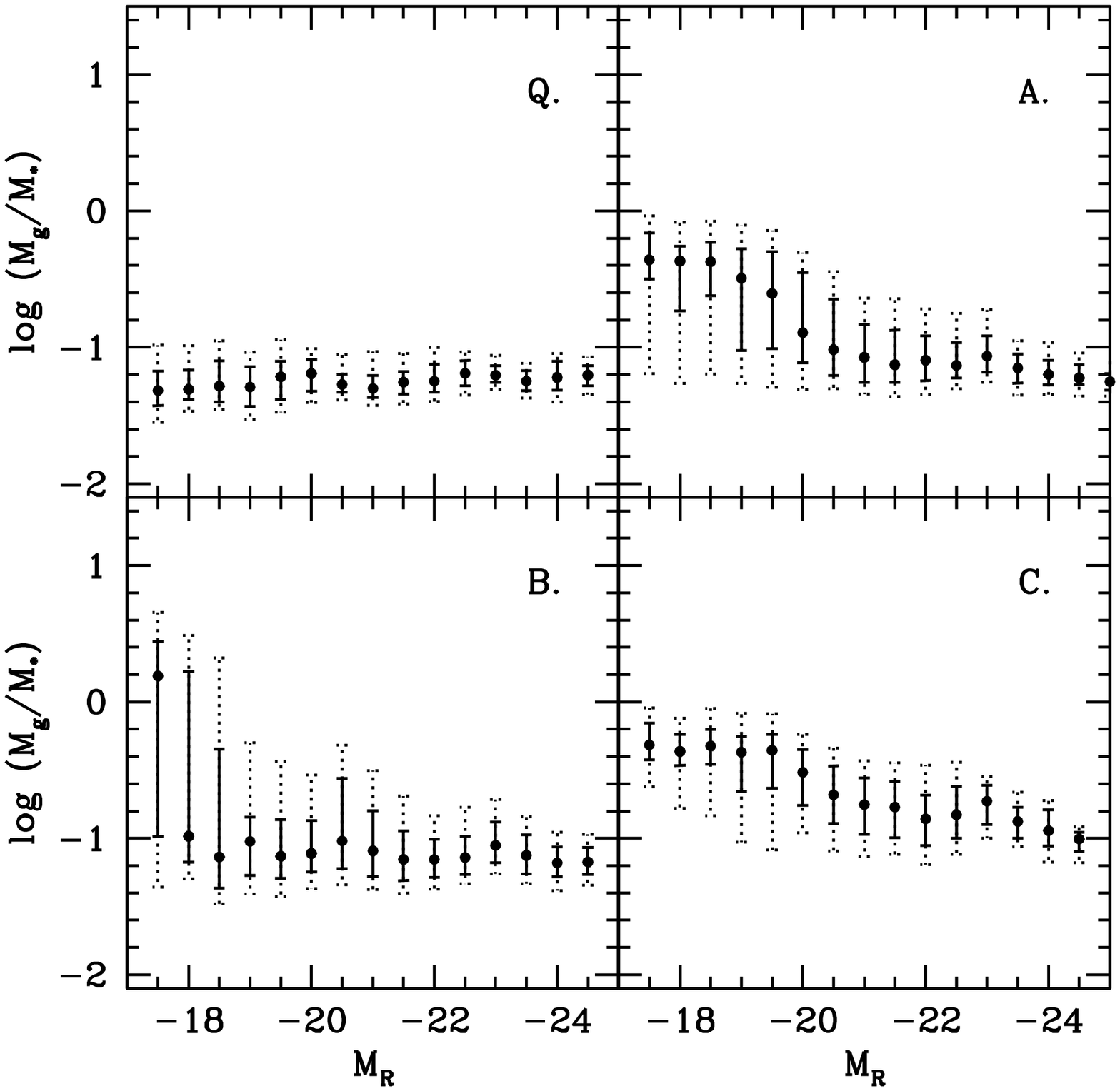}
}
\caption{\label{fig15}
\small
The ratio of gas mass to stellar mass is plotted as a function of               
$R$-band absolute magnitude 
for galaxies with $D(4000)<1.6$. }
\end {figure}

\section{Discussion and summary}
The use of observations at high redshift  
to reconstruct the formation histories of the
galaxies we see today has been a important goal of modern observational cosmology. 
In practice, this has turned out to be quite difficult. 
High-redshift galaxies differ from local galaxies in size, mass and morphology,
and there is no unambiguous way to identify the distant progenitors of normal spiral and
elliptical galaxies in the local Universe.
The use of stellar colour-magnitude diagrams to
unravel the past star formation histories of galaxies in the Local Group
has proved remarkably successful, but the extension of this technique to more
distant galaxies will not be possible for many years.

The analysis of galaxy populations provides another way of constraining {\em average} star
formation histories. To this end, analyses of 
the H$\delta$ distributions of cluster
galaxies have been carried out  by Barger et al (1996) 
and Poggianti et al (1999), but so far
there have been no similar studies of field galaxies. 
The key assumption is that all galaxies 
in the sample can
be regarded as Monte Carlo realizations of the same set of underlying physical
processes. The list of processes becomes extremely large if one
considers the entire population of field galaxies.  However, by judiciously choosing
subsamples, such as splitting galaxies by luminosity or environment, one is
able to test the basic assumptions of the model.
Another important consideration is that the completeness limits on
the survey must be well understood. If, for example, a galaxy forming stars with low   
efficiency falls below the surface brightness detection limits of the survey
and this effect is not taken into account, erroneous
conclusions will be obtained.         

In this paper, we have focused on the spectral rather than the photometric properties
of galaxy populations. As we have emphasized, the advantage of using spectral
indicators such as $D(4000)$ or H$\delta$  over broadband colours, is that 
they are sensitive to stellar age, rather than to some combination of age, metallicity
and extinction. The main disadvantage of using spectra in the analysis of 
future large surveys such as SDSS or 2dF is the problem of aperture bias 
(Kochanek, Pahre \& Falco 2000). This may be overcome to some extent by
selecting only galaxies small enough in angular diameter so that a significant fraction
of their total light has entered the fiber.

Finally, it would be interesting to explore other signatures of starbursts                      
in galaxies and see how well they correlate
with spectral properties. For example, it is known that a significant
fraction of the star formation in local starburst galaxies occurs via the
creation of super star clusters (SSCs), each with hundreds to thousands of OB
stars concentrated within a few parsecs (Whitmore et al 1999).
At least some SSCs are gravitationally bound and their masses suggest they may be
globular clusters in their infancy. If clustered star formation is a generic feature
of the high-efficiency `triggered' mode of star formation, then studying
these systems may provide another way of probing the importance of bursts in
the star formation histories of nearby galaxies.

In summary,
we have explored a model in which star formation occurs in two modes: 1) a
low-efficiency continuous  mode; and 2) a high-efficiency mode  
triggered by an interaction with a satellite more than 1\% of the primary's mass.
Merger/interaction rates are specified using the predictions of
hierarchical galaxy formation models.  
With these assumptions, the star formation histories of low-mass galaxies
are characterized by intermittent bursts of formation separated
by quiescent periods lasting several Gyr; massive field galaxies are perturbed
on timescales of several hundred million years and have more continuous 
star formation histories.  
We have explored the distribution of spectral indicators such as $D(4000)$ and  
H$\delta_A$ 
by combining the models with the latest version of the  Bruzual \& Charlot (1993) 
spectral synthesis code. In models where stars form only in the continuous mode,
the  star formation rate, H$\delta$ equivalent width and gas fraction distributions of
young galaxies with  $D(4000) < 1.6$ 
are narrowly peaked and do not depend on
luminosity. In the burst models, the distributions have much larger
dispersion and depend strongly on galaxy luminosity. 
The widths and shapes of the distributions place strong constraints on
the relative efficiency of the burst mode, as well as the typical duration of
the bursts.  We conclude that the analysis of spectral indicators
in future large redshift surveys will provide important new constraints on the nature of
the star formation histories of local galaxies.\\

\vspace{0.8cm}

\large
{\bf Acknowledgments}\\
\normalsize
We thank Jay Gallagher for helpful discussions and for convincing us to carry out this
project, and Ann Zabludoff for providing
the table of spectral indices for the LCRS survey.

\pagebreak 
\Large
\begin {center} {\bf References} \\
\end {center}
\normalsize
\parindent -7mm  
\parskip 3mm

Balogh, M.L., Morris, S.L., Yee, H.K.C., Carlberg, R.G. \& Ellingson, E., 1999, ApJ, 527, 54

Barger, A.J., Aragon-Salamanca, A., Ellis, R.S., Couch, W.J., Smail, I. \& Sharples, R.M., 1996, MNRAS,
279, 1

Barnes, D.G., Staveley-Smith, L., de Blok, W.J.G., Osterloo, T., Stewart, I.M.,
Wright, A.E., Banks, G.D., Bhathal, R. et al, 2000, MNRAS in press 

Barnes, J.E. \& Hernquist, L.E., 1991, ApJ, 370, 65

Borne, K.D., Bushouse, H., Lucas, R.A. \& Colina, L., 2000, ApJ, 529, 77

Bressan, A., Chiosi, C. \& Tantalo, R., 1996, A\&A, 311, 425

Bruzual, A.G., 1983, ApJ, 273, 105

Bruzual, A.G. \& Charlot, S., 1993, ApJ, 405, 538

Byrd, G.G. \& Howard, S., 1992, AJ, 103, 1089

Calzetti, D., Kinney, A.L. \& Storchi-Bergmann, T., 1994, ApJ, 429, 582

Charlot, S. \& Fall, S.M., 2000, ApJ, 539, 718

Charlot, S. \& Longhetti, M., 2001, MNRAS in press (astro-ph/0101097)

Cohen, J.G., 1976, ApJ, 203, 587

Cole, S., Lacey, C.G., Baugh, C.M. \& Frenk, C.S., 2000, MNRAS, 319, 168

Couch, W.J., Balogh, M.L., Bower, R.G., Smail, I., Glazebrook, K., \& Taylor, M., 2001,
ApJ in press (astro-ph/0010505)

Cowie, L.L., Songaila, A., Hu, E.M. \& Cohen, J.G., 1996, ApJ, 112, 839

Dalcanton, J.J., Spergel, D.N. \& Summers, F.J., 1997, ApJ, 482, 659

de Blok, E., van der Hulst, T. \& McGaugh, S.S., 1996, MNRAS, 283, 18

Ellis, R.S., 1997, ARA\&A, 35, 389

Dolphin, A.E., 2000, MNRAS, 313, 281

Driver, S.P., 1999, ApJ, 526, 69

Fanelli, M.N., O'Connell, R.W. \& Thuan, T.X., 1988, ApJ, 334, 665

Folkes, S., Ronen, S., Price, I., Lahav, O., Colless, M., Maddox, S.,
Deeley, K., Glazebrook, K. et al, 1999, MNRAS, 308, 459

Gallagher, J.S., Hunter, D.A. \& Bushouse, H, 1989, AJ, 97, 700

Gavazzi, G. \& Scodeggio, M., 1996, A\&A, 312,29

Gavazzi, G., Catinella, B., Carrasco, L., Boselli, A. \& Contursi, A., 1998, AJ, 115, 1745

Glazebrook, K., Abraham, R., Santiago, B., Ellis, R.S. \& Griffiths, R., 1998, MNRAS, 297, 885

Gunn, J.E. \& Gott, J.R., 1972, ApJ, 176, 1

Hernandez,X., Valls-Gabaud, D. \& Gilmore, G., 2000, MNRAS, 316, 605

Hernquist, L. \& Mihos, C.J., 1995, ApJ, 448, 41

Jansen, R.A., Franx, M. \& Fabricant, D., 2000, ApJ in press (astro-ph/0012485)

Kauffmann, G. \& White, S.D.M., 1993, MNRAS, 261, 921

Kauffmann, G., White, S.D.M. \& Guiderdoni, B., 1993, MNRAS, 264, 201

Kauffmann, G. \& Charlot, S., 1998, MNRAS, 294, 705

Kauffmann, G., Colberg, J.G., Diaferio, A. \& White, S.D.M., 1999b, MNRAS, 307, 529

Kennicutt, R.C., 1983, AJ, 88, 483

Kennicutt, R.C., 1983b, ApJ, 272, 54

Kennicutt, R.C., 1983c, A\&A, 120, 219

Kennicutt, R.C., 1992, ApJ, 388, 310

Kennicutt, R.C., 1998, ARA\&A, 36, 189 

Kennicutt, R.C., 1999, ApJ, 525, 1165

Kennicutt, R.C. \& Kent, S.M., 1983, AJ, 88, 1094

Kochanek, C.S., Pahre, M.A. \& falco, E.E., 2000, ApJ submitted (astro-ph/0011458)

Kolatt, T.S., Bullock, J.S., Dekel, A., Primack, J.R., Sigad, Y., Kravtsov, A.V. \&
Klypin, A.A., 2000, MNRAS submitted (astro-ph/0010222)

Lacey, C. \& Silk, J., 1991, ApJ, 381, 14

Lacey, C. \& Cole. S., 1993, MNRAS, 262, 627

Lemson, G. \& Kauffmann, G., 1999, MNRAS, 302, 111

Lin, D.N.C., Jones, B.F. \& Klemola, A.R., 1995, ApJ, 439, 652

Loveday, J., Tresse, L. \& Maddox, S., 1999, MNRAS, 310, 281

Matthews, L.D., Gallagher, J.S. \& van Driel, W., 1999, AJ, 118, 2751

McGaugh, S.S. \& Bothun, G.D., 1994, ApJ, 107, 530

Mihos, J.C. \& Hernquist, L., 1994, ApJ, 437, 611

Mihos, J.C. \& Hernquist, L., 1996, ApJ, 464, 641

Mo, H.J., Mao, S. \& White, S.D.M., 1998, MNRAS, 295, 319

Mo, H.J., McGaugh, S.S. \& Bothun, G.D., 1994, MNRAS, 267, 129

Moore, B., Katz, N., Lake, G., Dressler, A. \& Oemler, A., Nature, 379, 613

Navarro, J.F., Frenk, C.S. \& White, S.D.M., 1995, MNRAS, 275, 56  

Olsen, K.A.G., 2001, AJ, in press (astro-ph/9902031)

Poggianti, B.M., Smail, I., Dressler, A., Couch, W.J., Barger, A.J.,
Butcher, H., Ellis, R.S. \& Oemler, A., 1999, ApJ, 518, 576

Press, W.H. \& Schechter, P. 1974, ApJ, 187, 425

Rocha-Pinto, H.J., Maciel, W.J., Scalo, J. \& Flynn, C., 2000, ApJ, 531, 115  

Roukema, B.F., Peterson, B.A., Quinn, P.J., Rocca-Volmerange, B., 1997, MNRAS, 292, 835

Shectman,S.A., Landy, S.D., Oemler, A., Tucker, D.L., Lin, H., Kirshner, R.P. \&
Schechter, P.L., 1996, ApJ, 470, 172 

Somerville, R.S. \& Kolatt, T.S., 1999, MNRAS, 305, 1

Somerville, R.S. \& Primack, J.R., 2000, MNRAS, 310, 1087

Somerville, R.S., Primack, J.R. \& Faber, S.M., 2001, MNRAS, 320, 504 (SPF)         

Springel, V. , White, S.D.M., Tormen, G. \& Kauffmann, G., 2001, 
MNRAS submitted (astro-ph/0012055)

Sullivan, M., Treyer, M.A., Ellis, R.S., Bridges, T.J., Milliard, B. \& Donas, J., 2000,
MNRAS, 312, 442

Tolstoy, E., 2000, preprint (astro-ph/0010028)

Tormen, G., Diaferio, A. \& Syer, D., 1998, MNRAS, 299, 728

van den Hoek, L.B., de Blok, W.J.G., van der Hulst, J.M. \& de Jong, T., 2000, A\&A, 357, 397

van Kampen, E., Jiminez, R. \& Peacock, J.A., 1999, MNRAS, 310, 43

van Zee, L., Haynes, M.P. \& Salzer, J.J., 1997, AJ, 114, 2479

Wang, B. \& Heckman, T.M., 1996, ApJ, 457, 645

Whitmore, B.C., Zhang, Q., Leitherer, C., Fall, S.M.,
Schweizer, F. \& Miller, B.W., 1999, AJ, 118, 1551

Worthey, G., Faber, S.M., Gonzalez, J.J. \& Burstein, D., 1994, ApJS, 94, 687

Worthey, G. \& Ottaviani, D.L., 1997, ApJS, 111, 377

Yee, H,K,C., Ellingson, E. \& Carlberg, R.G., 1996, ApJS, 102, 269

York, D.G., Adelman, J., Anderson J.E., Anderson, S.F., Annis, J., Bahcall, N.A., Bakken, J.A.,
Barkhouser, R. et al, 2000, AJ, 120, 1579

Zabludoff, A.I., Zaritsky, D., Lin, H., Tucker, D., Hashimoto, Y., Shectman, S.A.,
Oemler, A. \& Kirshner, R.P, 1996, ApJ, 466, 104

\pagebreak 
\vspace {1.5cm}
\normalsize
\parindent 7mm  
\parskip 8mm

{\bf Table 1:} The parameters for the models                                                
\vspace {0.3cm}

\begin {tabular} {lcc}
 & $\alpha_b/\alpha_q$ & $t_b$ (Gyr) \\                                                                  
 Model Q & 0 & 1  \\
 Model A & 10 & 1  \\
 Model B & 100 & 1  \\
 Model C & 10 & 0.4  \\
\end {tabular}

\end {document}